\documentclass[12pt,a4paper]{article}
\usepackage{graphics}
\begin{document}
\title{Polarization catastrophe in doped cuprates and metal-ammonia solutions: an analogy.}
\author{P. Qu\'emerais$^1$, J.-L. Raimbault$^2$ and S. Fratini$^1$ \\
\small $^1$ Laboratoire d'Etudes des Propri\'et\'es Electroniques des Solides, CNRS, \\
\small avenue de martyrs, BP166, 38042, Grenoble Cedex 9. \\
\small $^2$ Laboratoire de Physique et Technologie des Plasmas, Ecole Polytechnique, \\
\small  91128, Palaiseau, Cedex}
\date{}

\maketitle
\begin{abstract}
 On doping polar dielectrics, such as the cuprates or liquid ammonia,
the long range polarization leads to the formation of bound states (polarons or
solvated electrons). However, the exact role of such entities in the
metal-insulator transition (MIT) still remains  unclear. We suggest
that the driving mechanism of the MIT is a polarization catastrophe
that occurs due to their unscreened Coulomb interaction.
This phenomenon is associated to a negative static dielectric
constant, which could be the origin of both the superconducting
transition in the cuprates --- where the doping ions are frozen in the
lattice structure --- and the phase separation observed in liquid
metal-ammonia solutions --- where the counter-ions are mobile. 
\end{abstract}

\thispagestyle{empty}
\pagestyle{empty}

\vspace{.5cm}

The aim of this paper is to propose an analogy between high-$T_c$ cuprate
superconductors (HTSC) and metal-ammonia solutions (MAS). The analogy
starts with the simple observation that for both systems, the MIT is
accompanied with a special instability. For cuprates, there is an
\textit{insulator-to-superconductor transition}, while for metal-ammonia
solutions the MIT is related with \textit{phase separation}. 
In recent works we have studied the melting of a Wigner crystal of
polarons as a function of both temperature and doping density (the
full details are presented in ref. [1]). Starting from the insulating
phase at low density, an insulator-to-metal transition takes place
upon increasing the density. There are several important results to
underline: (i) when the electron-phonon coupling is strong enough, the
melting toward a polaron liquid is not possible; (ii) rather, the
melting of the polaron Wigner crystal is driven by a polarization
catastrophe leading to \textit{polaron dissociation} above the MIT; (iii) at
the critical density, the \textit{static dielectric constant becomes
negative}. The last point indicates that the existence of a mixed phase
of free electrons coexisting with localized polarons above the
polarization catastrophe could lead to a superconducting instability
[2]. The glue for the superconductivity in this scenario would be due
to the vibrations of the residual localized electrons inside their
polarization potential-wells, which are able to overscreen the Coulomb
interaction between mobile electrons and induce their pairing.  
\begin{figure}[htbp]
  \begin{center}
    \includegraphics{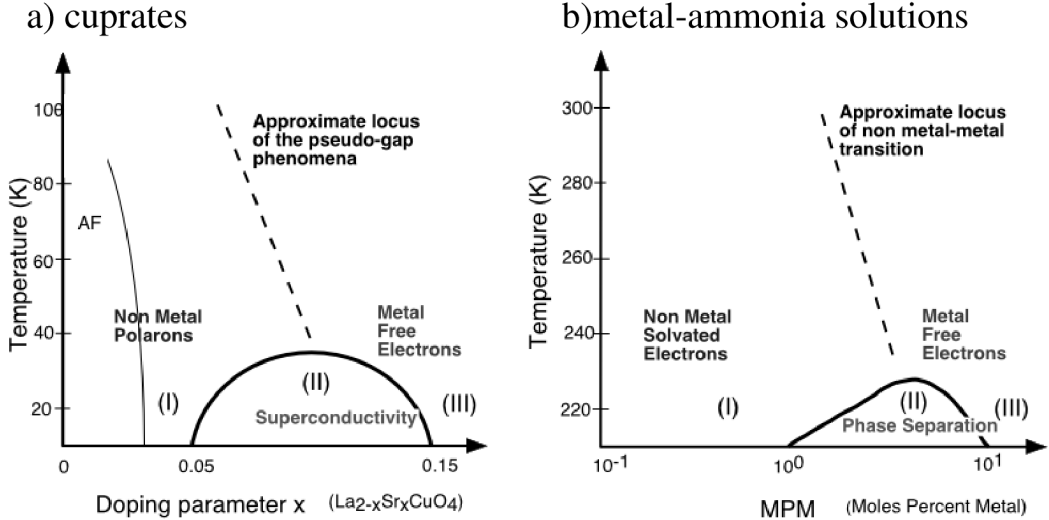}
    \caption{\footnotesize{Comparison between the  phase diagrams of a) doped
      cuprates (sketched), and b) metal ammonia solutions (from
      [4]). MPM corresponds to the molar fraction of metal. For both
      systems, the non metal -to-metal transition occurs for carrier
      densities close to $10^{20}-10^{21}$ per $cm^3$. The region (I) refers to
      the polaronic/solvated electron phase, region (III) to the
      normal metallic phase, while region (II) corresponds to a
      possible mixed state of polarons and free electrons.}} 
    \label{fig:phasediag}
  \end{center}
\end{figure}

The generic phase diagram of the doped cuprates is sketched on Fig.1a,
and can be roughly divided in three different regions. (I) is the low
density regime where all the doping charges form polarons, and
interact through the unscreened long-range Coulomb interactions. (III)
is the region where the formation of dielectric polarons is completely
prevented due to the screening of the interactions. There is no more
superconductivity at such high densities, and the system tends to an
ordinary metal. The straight line represents the onset of the
pseudo-gap phenomenon. This  can be extended down to zero temperature,
well inside the superconducting region, by applying a strong magnetic
field [3], indicating that the insulator-to-metal crossover takes
place in the region (II) around optimal doping. In our picture, the
region (II) corresponds to a mixed phase of free electrons and
localized polarons occurring above the polarization catastrophe. 

The phase diagram of liquid metal-ammonia solutions is sketched on
Fig.1b. As for the cuprates, there are three well defined regions. On
doping pure ammonia with alkali metals (e.g. with Na), the metal atoms
dissociate yielding two species in solution: the electrons, and the
metal ions that are solvated by the ammonia molecules. At low density
(I), the electrons are also solvated by ammonia molecules yielding
entities --- the solvated electrons --- which are equivalent to Fr\"ohlich
polarons in oxides. A simple model for solvated electrons was first
proposed by Jortner [5] and confirmed later by path integral
calculations [6] and numerical (molecular dynamics) simulations
[7]. The short range repulsion between the electron and the ammonia
molecules creates a small cavity (of radius about $3$ \AA) in which
the   
electron localizes (see Fig.2). The attractive forces giving rise to 
localization
come from the long range polarization due to the
orientational degrees of freedom of the ammonia
molecules. Equivalently, the formation of (Fr\"ohlich) polarons in
oxides comes from the polarization field created by the longitudinal
optical phonons of the underlying lattice. In both cases, the
long-range asymptotic potential felt by the electrons is Coulombic:
$V(r)\sim e/\tilde{\varepsilon} r$, with $\tilde{\varepsilon}^{-1}=
\varepsilon_\infty^{-1} -\varepsilon_s^{-1}$. Here
$\varepsilon_\infty$ is the optical dielectric constant ($\approx 4$ for
undoped cuprates, and $\approx 2$ for pure ammonia), determined by the response
of the ion cores, while $\varepsilon_s$ is the static dielectric 
constant ($\approx 30$ for
undoped cuprates, and $\approx 22$ for ammonia). 
The difference between $\varepsilon_\infty$  and $\varepsilon_s$
originates from the polarizability related to the lattice
deformation (in the cuprates), or the orientation of ammonia molecules
(which carry a permanent dipole moment), and indicates that both
systems are strongly polarizable. 

When the coupling of electrons to the polarization degrees of freedom
is strong, the potential-well which surrounds each electron has at
least two localized levels: the ground-state (1s) and one excited
state (2p). The transition energy is given by $\hbar
\omega_0=E_{2p}-E_{1s}$, and can be
measured by optical experiments. In the cuprates, 
$\omega_0\approx0.1-0.15 eV$ [8],
while in metal-amonia solutions $\omega_0\approx0.8-0.9 eV$ [9]. 
The polarization 
vibrates at a much lower frequency, and therefore is not able to
\textit{screen} such transition. 

\begin{figure}[htpb]
  \begin{center}    
\resizebox{13cm}{!}{\includegraphics{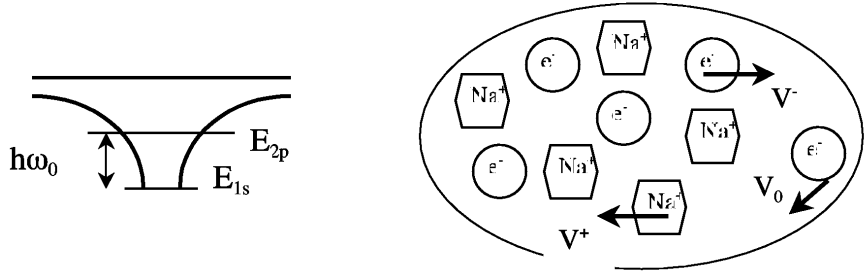}}
    \caption{\footnotesize{Left: Jortner's model [5]; the electron is 
localized inside a cavity, to form a solvated   
electron. The lowest energy levels in the potential-well are 
$E_{1s}$ (ground state) and $E_{1p}$ (excited state).
The transition energy is $\hbar \omega_0=E_{1p}-E_{1s}$.
Right: The three fluid model for metal-ammonia solutions consists of:
the solvated metal ions (Na+) with velocity $V^+$, the neutral cavities
(circles) with velocity $V_0$, and the electrons (e-) with velocity $V^-$,
which  are linked to the cavities by springs of strength $K$.}} 
    \label{fig:model}
  \end{center}
\end{figure}
Starting from the low density limit, we can calculate the dielectric
constant as we did for the polaron Wigner crystal in references
[1,2]. Let us stress that microscopic models for one (or two) solvated
electrons exist [5,6,7], but the many solvated-electrons problem has
been poorly studied to our knowledge.  
We propose here a simple hydrodynamic (and classical) model able to
incorporate some important aspects of the many body problem. For the
sake of simplicity, we only consider three kinds of particles in
solution [10]: the solvated metal ions with mass $M^+$ and charge $+e$, the
neutral cavities with mass $M_0$ (we take $M_0=M^+\equiv M$), and the electrons
with mass $m_e$ and charge $-e$. Each electron is linked to a neutral
cavity through a harmonic oscillator with spring constant
$K=m_e\omega_0^2$ 
[11] (see figure 2 ). The basic equations of motion are: 
\begin{eqnarray}
n^- m_e \dot{\mathbf{V}}^-&=&-n^-e\mathbf{E}_{loc}-n^-m_e\omega^2
(\mathbf{R}^--\mathbf{R}_0)-n^-m_e\Gamma_e\mathbf{V}^-  
\nonumber\\
n^+M \dot{\mathbf{V}}^+&=&+n^+e \mathbf{E}_{loc}-k_BT\nabla n^+
-n^+M\Gamma \mathbf{V}^+
  \label{eq:1}\\
n_0M \dot{\mathbf{V}}_0&=&+n_0 \eta M \omega_0^2 
(\mathbf{R}^--\mathbf{R}_0) -k_BT\nabla n_0-n_0M\Gamma\mathbf{V}^+ .\nonumber
\end{eqnarray}
where $\eta=m_e/M$ is the mass ratio ($\eta\ll 1$), $\Gamma_e$
and $\Gamma$ are respectively the
damping constants of the electrons and of the ions (and cavities). The
terms containing gradients of the density correspond to the pressure
forces in the particle fluid. To evaluate the dielectric constant, we use the
expression of the current
density: 
\begin{equation}
  \label{eq:2}
  \mathbf{j}(\mathbf{k},\omega)=
i\omega[\varepsilon_\infty-\varepsilon(\mathbf{k},\omega)]
\mathbf{E}(\mathbf{k},\omega)=
-e\bar{n}[\mathbf{V}^-(\mathbf{k},\omega)-\mathbf{V}^+(\mathbf{k},\omega)]
\end{equation}
($\bar{n}$ is the average electron density). 
The local field, which takes into account in mean field the effect of all other fluid particles, is related to the macroscopic field in the medium by
the Lorentz formula:  
\begin{equation}
  \label{eq:3}
 \mathbf{E}_{loc}=\frac{\varepsilon(\mathbf{k},\omega)+2\varepsilon_\infty}
{3\varepsilon_\infty}\mathbf{E}(\mathbf{k},\omega)  .
\end{equation}
Using the standard conservation
equation for mass and expanding in powers of the mass ratio $\eta$, one gets an approximate
expression for the real part of the dielectric constant (details of
the calculations will be published elsewhere):  
\begin{equation}
  \label{eq:4}
Re[\varepsilon(\mathbf{k},\omega)]\simeq \varepsilon_\infty+
\frac{\omega_p^2[-\omega^2+(k_BT/m_e)k^2+
(\omega_0^2-\omega_p^2/3\varepsilon_\infty)]}
{[(k_BT/m_e)k^2+(\omega_0^2-\omega_p^2/3\varepsilon_\infty)-\omega^2]^2
+\Gamma_e^2 \omega^2}\; \left\lbrace 1+\eta C_1(\mathbf{k},\omega)+O(\eta^2)
\right\rbrace
.  
\end{equation}
The coefficients ($C_i$) of the expansion are bound quantities, so that 
when the density satisfies $\omega_p^2/3\varepsilon_\infty \to
\omega_0^2$, the uniform ($k=0$)
dielectric constant diverges as $1/\Gamma_e^2$ when $\omega$ tends to
zero. The above
condition defines the critical density for the  \textit{polarization
catastrophe}. At the same time, \textit{the static dielectric constant becomes
negative} [12]. These results generalize to metal-ammonia solutions
what we already obtained for the Wigner crystal of polarons [1,2]. In
both cases, a transfer of spectral weight from high to low frequency
is expected when approaching the instability, in
agreement with the "color change" observed in both metal-ammonia
solutions and high-$T_c$ superconductors [13]. 

Experiments indicate that the electrons added by doping beyond this
critical density do not form bound states, but rather remain free. The
properties of the resulting mixed phase now depend on the behavior of
both the electrons, and of the compensating positive charges which
also feel a negative dielectric constant. If the latter are free to
move, as is the case in liquids, they naturally tend to produce phase
separation. On the contrary, if they are frozen in a structure as in
the doped cuprates, phase separation is prevented, and the
overscreening due to the negative sign of the dielectric constant can
induce pairing of the free electrons [14]. 

The present discussion remains of course qualitative. It presents some
conjectures about what happens beyond the instability of the
insulating phase of a polarizable medium. The real situation is
certainly more complicated than expected in the present scenario. For
example, multipolaronic states arranged along strings could appear as
proposed by Kusmartsev [15]. However, we wish to emphasize the 
following points: (i) the role of polarons in the MIT has been
underestimated up to now; (ii) the long range Coulomb forces may be
fundamental for the understanding of systems such as the
superconducting cuprates (see e.g. ref.[16]) or the metal-ammonia
solutions. 


\begin{thebibliography}{99}
\bibitem{[10]} There are other species in solution such as  ``bipolarons'' at
intermediate concentrations, and   
certainly also complexes of solvated metal ions and solvated electrons.
\bibitem{[11]} Rigourously, the solvated electron is the problem of a quantum
particle interacting with a system of  
classical liquid particles (see [6]). The quantum transition of the
electron from (1s) to (2p) state is  
here approximated by a classical harmonic oscillator with frequency $\omega_0$.
\bibitem{[14]} In the few cases where the doping ions are mobile enough, as in
La$_2$CuO$_{4+y}$, phase separation prevails over superconductivity.
\end{thebibliography}
\end{document}